# Introduction of Empirical Topology in Construction of Relationship Networks of Informative Objects

Hesam T. Dashti[1], Mary E. Kloc[2], Tiago Simas[3], Rita A. Ribeiro[3], Amir H. Assadi[1*]

[1] Department of Mathematics, University of Wisconsin, USA
[2] Department of Applied Mathematics and Computer Science, Weizmann Institute of Science, Israel
[3] CA3, Uninova, Campus New University of Lisbon/FC, Portugal

dashti@wisc.edu, mary-elizabeth.kloc@weizmann.ac.il, tms@uninova.pt, rar@uninova.pt, ahassadi@wisc.edu

**Abstract.** Understanding the structure of relationships between objects in a given database is one of the most important problems in the field of data mining. The structure can be defined for a set of single objects (clustering) or a set of groups of objects (network mapping). We propose a method for discovering relationships between individuals (single or groups) that is based on what we call the *empirical topology*, a system-theoretic measure of functional proximity. To illustrate the suitability and efficiency of the method, we apply it to an astronomical data base.

## 1 Introduction

Determining the interrelationships of objects belonging to either a given data set (clustering) or networks (network mapping) is a central problem in many fields, including data mining (web mining, unsupervised classification) [1][2], computational biology (quantitative trait loci, analysis of gene expression profiles [3] and protein-protein interaction networks [7]), and computational astronomy (clustering profiles of variable stars, determining types of planets by luminosity clustering [4][5][6]). This motivates research groups in diverse areas to develop new practical methodologies.

The process of clustering or network mapping requires a similarity measure for grouping individuals; this measurement should be defined on the space of attributes of the individuals. The manner in which databases are constructed and individuals are grouped is strongly dependent on the aim of the researchers; for example, for finding similar types of stars in computational astronomy [8][9], the individuals could be the observed objects in the sky and the attributes could be the parameters that describe the individuals. Alternatively, for determining the most significant parameters (a feature reduction problem, in general), the individuals could be the parameters and the attributes could be the different types of stars. Hence, the similarity measure is strongly dependent on the types and roles of individuals and attributes in the given problem.

[*] The author gratefully acknowledges partial financial support through NSF Grants DMS-0923296 and DBI-0621702.

When analyzing data sets pertaining to individuals and attributes, two main issues must be considered: the level of noise in the data, and the size of the data sets. In general, noise reduction methods are database-dependent; there is no practical noise reduction algorithm for all types of databases and all types of noise. Some algorithms [10][11] focus on specific kinds of databases (and associated noisy data), while others incorporate some flexibility in order to deal with different kinds of noisy data. Also, there are now many improved clustering algorithms that are suitable for use on very large data sets, as designing algorithms with lower time complexity has been a central focus in computer science from the beginning [12].

In this paper, we propose a new method for Discovering Relationship Networks based on Empirical Topology (*ET-DRN*). The ET-DRN method is a procedure for grouping individuals (in both clustering and network-mapping problems) that can deal with many types of noisy data. It is a hierarchical clustering procedure that composed of three different algorithms: 1) the Euler algorithm is used to assign the individuals to groups, 2) a Genetic algorithm is used to increase density (as measured using the Shannon entropy) of individuals inside group and, 3) the dissimilarity of the groups is determined using the Kullback-Leibler (KL) divergence. One can easily generalize and adapt this procedure by replacing KL-divergence by a different information-theoretic variant. The ET-DRN algorithm clusters a given data set by dividing it into groups of objects where elements of each group have minimum possible entropy and the groups have maximum distance between them.

We use an astronomy database to evaluate the suitability of the ET-DRN; in this case, the aim is to discover relationships between stars observed by the OGLE project [13]. The objects (individuals) are observed in a special period of time and positions of the objects rather than the earth construct a time series in its dynamic space. Based on the time series and the color of the objects in the sky, 13 attributes are determined for each object, as shown in Table 1. A comprehensive description of the attributes can be found in [14]. As described in Table 1, the attributes are basically calculations of different layers (e.g. first frequency of first harmonic, first frequency second harmonic), so normalization of the data is necessary to obtain a single feature space. The ET-DRN algorithm uses heuristic algorithms that treat the individuals as a discrete collection of measurements and optimize various statistical criteria. It takes advantage of a system-theoretic measure of functional proximity (called *empirical topology*) and applies it to the computational astronomy case study by comparing the different observed objects with space-like snapshots (intervals) of the complex dynamical system. Each snapshot is determined by using the Euler method on one attribute $a_i$. The Shannon entropy is computed for attributes $a_{i+1}$ through $a_{13}$. Then ET-DRN rearranges the objects of each snapshot to reduce the snapshot's system entropy. Finally, hierarchical clustering is performed on attribute $a_{i+1}$.

The empirical topology imposes an organization of the astronomical observations that is analogous to the subdivision of a topological space into its connected components. This allows the data to be organized into a network with far fewer nodes

(snapshots), where each node encodes the measurements of a particular class of stars. Considering individuals in a high dimensional space and analyzing them as objects of a digital geometry allows us to construct a sensible mathematical structure (to which all of the approaches of computational geometry can be applied) where noisy data and outliers are clearly detectable. In a more general setting, digital geometries can be regarded as finer mathematical structures that are imposed on sets of points arising from discrete samples. *Neighborhood relationships* of collections of observations could be regarded as sets of points in a digital space.

The ET-DRN algorithm handles all objects in the database according to their positions in the space of features. As mentioned, ET-DRN is a hierarchical clustering algorithm that, in each step, reduces the entropy of the system to just one dimension of the feature space and finds intervals of similar objects in the corresponding feature space. Unlike other iterative hierarchical clustering algorithms, the ET-DRN algorithm iterates with attention to the distribution of objects in other dimensions, so in each step it is working on only a small part of the database, focusing on one of the features while paying attention to this feature's relationship with all other features. During the process of reducing the system entropy, the ET-DRN algorithm increases the distance of classes using the Kullback-Leibler divergence which incorporates the probability of connectivity of the objects to determine the more significant classes.

As mentioned before, we applied the ET-DRN algorithm to an astronomical data set, where Simas et. al. [9] had already performed comprehensive study on appropriateness and accuracy of several clustering algorithms to the same set. In their study, they examined the performance of the algorithms at hand based on testing six desirable features; but *none of the outcomes of six algorithms could produce satisfactory results on tests of all six features*. Here we show that beside precision of the ET-DRN algorithm, it satisfies all of the desired features, thus demonstrating a different level of performance that was not hitherto achievable. Therefore, the failure of previous algorithms on the proposed features in [9], is sufficient grounds that we need not compare our algorithm with the other ones, and just demonstrate the ET-DRN properties for satisfying those features.

The Methodology section describes the empirical topology and outlines each step of the algorithm in detail. Section 3 presents results using experimental data that demonstrate the suitability of the ET-DRN for this problem.

## 2 Contribution to technological innovation

Non-supervised clustering algorithms have proved to be of great value in the increasingly data-dependent technological advances. An illustration of this statement is in applications to new and improved medical instrumentation, especially in brain surgery as applied to devastating diseases such as Parkinson's Disease. One of the senior co-authors have two one US patent and another international patent for application of the unsupervised clustering of data formed from brain signals, such as spiking of neurons collected by multi-electrodes [18][19][20]. The specific applications to Deep Brain Stimulation (DBS) are particularly noteworthy, and the results to relieve severe symptoms in Parkinson's patients are truly remarkable and highly praised in literature. Unfortunately, as in all other complex diseases of the nervous system, the applicability of the DBS for Parkinsonian patients through DBS is limited to individual circumstances of the patient and types of the insult to the brain. Nonetheless, such algorithms provide much more precise tools for the neurologists and neurosurgeons to utilize advanced medical instrumentation to advantage.

The unsupervised algorithms in the ET-DRN class presented in this article also fall in that category, as shown in the previously referenced algorithms [18]. The advantages of the ET-DRN approach to design of biomedical data clustering algorithms are in versatility and applicability to massive data sets, thus opening the potential for on-line and real-time analysis of massive data sets that must use many more (neuronal signal) recording channels for improved biomedical applications.

## 3 Methodology

The ET-DRN algorithm considers objects as points in a digital geometry of the high-dimensional space of their attributes. The process of the hierarchical clustering to reduce the dimension of the space is done by determining snapshots (intervals) of similar objects on each dimension. In other words, on each level of attributes, ET-DRN determines the groups of objects and then finds subgroups on higher levels of attributes. The snapshots are determined by using the Euler method. The Euler method of interpolating a function is applied to a given set of sample points (e.g. two dimensional) of a function and uses no apriori knowledge of the function. It starts from first sample node (start-point) and defines a straight line from the start-point to another point (end-point) where the line connecting these two points satisfies the constraint on slope change. Iteratively, the end-point then becomes the start-point for the next step, and the method is repeated until the last sample point. The groups of points between the start-points and end-points show snapshots (intervals) of the points. These groups, upon definition of the employed constraints, can represent clusters of objects in a given data base, where each sample point is associated with two attributes of an object. The following pseudocode describes the implementation of the Euler method and the employed constraints on it.

```
program EulerConstrain (slope)
  {Assume 'fabs(X)' function returns absolute value of
X. input: FirstIndex, FirstValue, SecondIndex,
SecondValue };
  const   NeperNumber = 2.71;
  var     Power;
begin
    Power = fabs(FirstValue - SecondValue)
*fabs³(FirstIndex - SecondIndex);
    slope = (0.5)*(NeperNumber^Power -1);
end.
```

```
program EulerMethod (NumberClasses)
  {Assuming NumberClasses shows number of classes;
ObjectIndex traces on all objects in the given data
set; this program renews ClassID of all objects. Input:
DataSet};
  const  MaxYears = 10;
  var    ClassNum: Integer;
  begin
    For all ObjectIndex on DataSet
      begin
        if EulerConstrain(ObjectIndex.Index,
          ObjectIndex.Value, ObjectIndex->Next.Index,
          ObjectIndex->Next.Value) > Threshold
          begin
            NumberClasses = NumberClasses + 1;
          end.
        ObjectIndex->Next.ClassID
      end.
  end.
```

As mentioned, the ET-DRN algorithm gradually reduces the dimension of the space of objects using the Euler method. The ET-DRN considers the influence of all attributes on the arrangement of objects (sample points in the Euler method). In each step of the hierarchical clustering, the ET-DRN rearranges the objects of the intervals calculated in the previous step. In each interval, the optimal arrangement is that closest objects settle beside each other, where the measure of closeness of two objects is the $L_2$ norm of their unreduced attributes. In order to find the best arrangement, ET-DRN assumes that the objects in their digital geometry are cities of the Traveling Salesman Problem (TSP) in high-dimensional space [15]. In fact, the ET-DRN reduces the problem to the TSP and tackles the TSP using the Genetic Algorithm (GA). GA is notable heuristic approach; its convergence was proven by [16]. The topology of the employed GA is described in the comments of the following GA pseudocode:

```
program GA (NewArrangment)
  {Input: Dataset};
  const  MaxYears = 10;
  var    Population: two dimensional array of integer;
 begin
            Population[PopNum];//Population[i]represents
                              //an arrangement for
                              //elements of the Dataset

  for i:1..PopNum
    begin
            Fill(Population[i]);//the "Fill" function,
```

```
                        //fills Population[i]
                        //with a random arrangement
                        //in normal distribution.
     end.
  While (satisfaction)//will stop when min entropy of
                      //the system reaches to a
                      //predetermined value or iterates
                      //for a predetermined times.
     begin
             Entropy(Population);//for each arrangement
                                 //of the Population
                                 //calculates L2 distance
                                 //of objects, where
                                 //objects are distributed
                                 //in the space of
                                 //attributes.
             Evolution(Population, EPopulation);
             Entropy(EPopulation);
             RolledWheel(Population, EPopulation);//Based
                                 //on rolled wheel method
                                 //replace the Population
                                 //with a combination of
                                 //previous Population and
                                 //EPopulation.

     end.
```

The clustering process of ET-DRN is completely flexible and based on a selected threshold in the constraint of Euler method; this allows the clustering approach to find a wide range of numbers of clusters. Typically, this ability is mentioned as a benefit of clustering algorithms, where the algorithm does not need any apriori information about the number of clusters. However, finding an automated approach for monitoring the number of clusters is very important to avoid obtaining insignificant clusters. Cluster monitoring and the ability to merge insignificant classes in order to get a smaller number of meaningful clusters are considered in the ET-DRN algorithm by calculating the Amount of Weights of Relation (AWR) of the clusters.

The AWR of the clusters is the same as distance of clusters from each other and could also be employed for network mapping problems, since this value is a measure of the connectivity of the clusters. In the case of clustering problems, this value is employed to find close clusters and merge them in order to discover significant groups of objects. The ET-DRN calculates the AWR by applying an alternative to the standard Kullback-Leibler divergence formula that we call "Entropic Kullback-Leibler" [Formula.1] to avoid confusion with the established terminology. The Entropic Kullback-Leibler uses entropy values where the traditional Kullback-Leibler formula utilizes probability values. The entropy value for each class comes from result of the GA procedure from previous step.

$$\text{Divergence}(i, j) = \text{Entropy\_i} * \log(\text{Entropy\_i}/\text{Entropy\_j}) \qquad (1)$$

The Kullback-Leibler formula is not symmetric, so the matrix of the AWR values is an asymmetric matrix. One can look at this asymmetric matrix as a weighted directed graph, and any kind of clustering algorithm can be employed for the network mapping problem.

In clustering algorithms, one problem that appears during the process of merging classes is deciding, when two classes are close together, which class should be merged to another one. In the ET-DRN algorithm, this problem is solved by using the attitude of the AWR matrix, where attractions of two classes differ from one class to another: the class more attractions will be absorbed by another class. The following pseudocode shows the method for constructing the AWR matrix:

```
program AWR_Matrix (Matrix: in dimension of
NumberClasses by NumberClasses)

{for a given array "Entropy" that includes entropy
values of all classes}
   var   i: 1..NumberClasses; j:1..NumberClasses

begin
   Matrix[i][j] = Divergence(i, j);
end.
```

## 4  Experimental Results

To evaluate the ET-DRN algorithm, we started by applying it to a labeled astronomical database [13]. The precision of the algorithm was computed based on the previous labels. In both databases, objects are represented by 13 attributes (Table 1), a discussion of which can be found in [14]. The labeled database includes approximately 10,000 objects, categorized into 9 classes.

**Table 1.** Name and description of the employed attributes

| Name | Description |
|---|---|
| Log-f1 | log of the first frequency |
| Log-f2 | log of the second frequency |
| Log-af1h1-t | log amplitude first harmonic first frequency |
| Log-af1h2-t | log amplitude second harmonic first frequency |
| Log-af1h3-t | log amplitude third harmonic first frequency |
| Log-af1h4-t | log amplitude fourth harmonic first frequency |
| Log-af2h1-t | log amplitude first harmonic second frequency |
| Log-af2h2-t | log amplitude second harmonic second frequency |
| Log-crf10 | amplitude ratio between harmonics of the first frequency |
| Pdf12 | phase difference between harmonics of the first frequency |
| Varrat | variance ratio before and after first frequency subtraction |
| B-V | color index |
| V-I | color index |

The precision values were calculated using the well-known True Positive formula (Formula 2). In this formula, "True Positive" is the number of correctly clustered objects and "False Positive" is the number of objects that were clustered incorrectly. This value is computed for each class separately, as listed in Table 2.

$$\text{Precision} = \text{True Positive}/(\text{True Positive} + \text{False Positive}) \qquad (2)$$

**Table 2.** Precision values for each class.

| Class ID | Precision |
|---|---|
| Class 1 | 0.987053 |
| Class 2 | 0.993354 |
| Class 3 | 0.956856 |
| Class 4 | 0.690141 |
| Class 5 | 0.861775 |
| Class 6 | 0.7625 |
| Class 7 | 0.520392 |
| Class 8 | 0.641975 |
| Class 10 | 0.8610 |

## 5 Conclusion

In this paper, we introduced a novel algorithm for constructing relationship networks of data sets of individuals based on a system-theoretic measurement of *functional proximity* (called *empirical topology*). The method, called ET-DRN, uses heuristic algorithms and performs a hierarchical clustering process on the representative attributes of the individuals. The ET-DRN algorithm was applied to an astronomical database, and its calculated precision was very promising for the de novo clustering problem. In addition, this algorithm was able to discover new relationships within ambiguous and incomplete databases such as the astronomical database using its hierarchical clustering procedure. The ET-DRN algorithm also allows for the merging of classes based on the value of the Kullback-Leibler divergence, thus increasing its usability and making it independent of any apriori information about the data base. We believe this method can be applied to any kind of database. Since it would be easy to transform into a parallel processing platform, the ET-DRN algorithm can be applied to very large databases, such as biological databases and future astronomical databases from European Space Agency ($10^8$) [17].

**Acknowledgments.** We gratefully acknowledge Luis Maro Sarro for discussions that led to an improved understanding of the problem characteristics, and for providing the database and permitting the comparison of our results with his excellent work. Amir Assadi gratefully acknowledges partial financial support through NSF Grants DMS-0923296 and DBI-0621702.